\documentclass{article}
\usepackage{graphicx,A4wide}
\begin{document}
\title{PROPERTIES OF SOME RESONANCES OF THE $^{55}\mbox{Mn}(p,\gamma)^{56}\mbox{Fe}$ REACTION IN THE E$_p$ = 1.3 - 1.8 MeV REGION}
\author{J. Materna, J. Lipt\'{a}k, F. \v{S}t\v{e}rba, J. \v{S}t\v{e}rbov\'{a}, J. Vrzal}      
\maketitle

\begin{abstract}
Radiative decay of 21 resonances in the $^{55}$Mn(p,$\gamma$)$^{56}$Fe reaction was studied in the proton beam energy region E$_{p}$ = 1.3 -- 1.8 MeV. Branching of decay to many low lying bound states up to excitation energy E$_{x}$ $\sim$ 8 MeV was measured. Exact energy of all resonances has been established what pointed out that five of the resonances are very close doublets. For all studied resonances were determined their spin-parity charakteristics. Assignment of some resonances as isobaric analogues of the states in the $^{56}$Mn nucleus was discused and short note about energy systematics of isobaric analogue resonances was shown.
\end{abstract}

\section{Introduction}

The excitation functions for the (p,p'$\gamma$), (p,n$\gamma$) and (p,$\gamma$) output channels of the proton induced reactions on $^{55}$Mn nucleus studied by our group were published in the paper~\cite{Ster}. The yields of $\gamma$-transitions from the first excited states in final $^{55}$Mn, $^{55}$Fe and $^{56}$Fe nuclei measured in the proton beam energy region E$_{p}$~=~1.3~--~2~MeV exhibit many well expressed resonances. Further experimental work has been devoted to the total $\gamma$-ray spectrum of the radiative decay of 21 individual resonances and to properties of capture and some bound states in the $^{56}$Fe compound nucleus. Therefore the $\gamma$-ray spectra containing the primary as well as secondary transitions ("primaries" and "secondaries"), were registered in a wide $\gamma$-ray energy region E$_{\gamma}$~$\sim$~0.3~--~12~MeV.

Radiative decays of some studied resonances was already measured in~\cite{Frid}~--~\cite{Guo}, however except ref. ~\cite{Frid}, usually only stronger primaries were presented. More, because rather thick targets ($\Delta$E$_{p}$ = 5 -- 6 keV) used in ~\cite{Frid}~--~\cite{Kaz2}, some transitions might be attributed to different resonances. Nevertheless, all results are mutually in reasonable agreement.

In present work we have tried to identify rather all possible transitions, but this aim is complicated by a very high complexity of the experimental $\gamma$-ray spectra. Nevertheless, we hope that we have found all stronger primaries up to the levels with excitation energy E$_{x}$ $<$ 6 MeV. High accuracy of the energy of the observed transition together with exact knowledge of lowest bound states energies made it  possible substantialy appreciate energy of the resonant states.

Although bound states in the $^{56}$Fe nucleus have been also studied, present paper is dedicated to the resonances. For enormous extend of obtained material the bound states should be discused in separate publication.

\section{Experimental conditions and methods}

\begin{table}[h]
{\small \begin{tabular}{|l@{}|r|r|r|c@{}|r|r|}
\hline 
&&
\multicolumn{2}{|c|}{$E_r[kev]$}&
&
\multicolumn{2}{|c|}{$J^\pi$}
\\ \cline {3-4} \cline{6-7}
n. &
$E_p[keV]$&
our&
NDS$^\star$ &
r(p,$\gamma$) &
our &
NDS$^\star$ 
\\ \hline \hline
1&1344.05(29)&$^{1)}$11503.89(23)&11509(4)&45&$^{a)}$3$^{+}$IAS&3$^{+}$IAS\\ 
2&1440.95(20)&11599.06(9)&$^{*}$11598.75(19)&21&1$^{+}$IAS&1$^{+}$IAS\\ 
3&1455.39(27)&11613.24(20)&$^{*}$11613.03(19)&31&1$^{+}$IAS&1$^{+}$IAS\\
4&1480.66(25)&11638.06(18)&---& 40 &3($^{-}$)&\\       
5&1483.44(25)&11640.79(18)&---&-"- &3($^{-}$)&\\
6&1486.85(33)&11644.29(28)&---&-"- &3($^{-}$)&\\
7&1507.21(27)&11664.14(20)&---& 28 &(3$^{-}$)&\\
8&1521.37(32)&11678.04(26)&---& 69 &4$^{+}$IAS&\\
9&1524.10(22)&11680.72(13)&---&-"- &4$^{+}$IAS&\\
10&1531.84(20)&11688.33(9)&11688.6(4)&57&$^{b)}$4$^{+}$IAS&4$^{+}$IAS\\
11&1535.86(23)&$^{2)}$11692.27(14)&11694(4)&112&$^{c)}$2$^{+}$IAS+4$^{+}$&2$^{+}$IAS\\
12&1679.04(19)&11832.90(7)&11832(4)&39&$^{d)}$3$^{+}$IAS&4$^{+}$IAS\\
13&1687.24(24)&$^{3)}$11840.95(17)&11840(4)&40&3$^{+}$IAS&4$^{+}$IAS\\
14&1696.47(46)&11850.02(42)&11848(4)&28&(3$^{+}$IAS)&3$^{+}$IAS\\
15&1726.70(30)&$^{4)}$11879.71(24)&---&40&(5$^{+}$IAS&\\
16&1734.00(37)&11886.88(32)&---&-"- &(5$^{+}$IAS)&\\
17&1761.01(55)&11913.40(52)&11918(4)&21&(4$^{+}$)&4$^{+}$5$^{+}$IAS\\
18&1773.17(21)&11925.35(12)&- " -&53&3($^{-}$)&- " -\\ 
19&1796.07(22)&11947.84(13)&---&32&(4$^{-}$)&\\
20&1801.03(27)&$^{5)}$11952.71(20)&---&52&4$^{+}$&\\     
21&1806.62(25)&11958.14(16)&11958(4)&46&(3$^{+}$IAS)&3$^{+}$IAS\\
\hline
\end{tabular}}\\
{\tiny $^{a)-d)}$Assignment from angular distribution ~\cite{Trun}: $^{a)}$ J$^{\pi}$ = 3$^{+}$,$^{b)}$ J$^{\pi}$ = 4$^{+}$,$^{c)}$ J$^{\pi}$ = ?,$^{d)}$ J$^{\pi}$ = 3$^{+}$\\
$^{1)-5)}$Probably doublets wit energies:\begin{tabular}[t]{l}
$^{1)}$ 11503.31(14) + 11504.23(12)\\
$^{2)}$ 11691.58(16) + 11692.40(7)\\
$^{3)}$ 11840.83(11) + 11841.70(24)\\
$^{4)}$ 11879.26(23) + 11880.17(16)\\
$^{5)}$ 11952.31(19) + 11953.13(12)
\end{tabular}\\
$^{*}$ Values taken from ~\cite{Guo}
}
\caption{Measured resonances in $^{56}$Fe}
\end{table}

Experimental arrangement was described in paper~\cite{Ster}~and in more details in~\cite{Ninh}. The measurements were performed on the 2.5 MeV Van de Graaff accelerator of the Faculty of Mathematics and Physics of Charles University in Prague. Thin targets (proton energy loss about 2 -- 3 keV for proton beam energy E$_{p}$~=~1.2~MeV) were prepared by vacuum evaporation of pure $^{55}$Mn onto 0.1 mm thick gold backing. More than 60 measurements were realised for 21 ``main'' resonances in the proton beam energy region E$_{p}$ = 1.3 -- 1.8 MeV, but five of these resonances are probably multiplets. Total charge collected onto the target for individual measurements was 0.1 -- 0.8 C in dependence on the resonance's strength. Corresponding running time was about 2 -- 12 hours. Fulfilment of resonant condition was controlled by keeping the rate of strongest $\gamma$-ray transition in final $^{56}$Fe nucleus (E$_{\gamma}$ = 847 keV) at maximum during all running time.

All resonances for which the $\gamma$-ray spectra were measured are collected in Tab.~1. In first and second columns is the proton beam energy in the laboratory system (all over assigned as E$_{p}$) and energy E$_{r}$ of the resonant state (capture state), respectively (see part 4.1. of this paper). In third column is the relative strengthness of the resonances in the (p,$\gamma$) reaction taken from Ref. ~\cite{Ster}. The literature information are taken from Nuclear Data Sheets (1999)~\cite{NDS}~and from paper published by Guo et al. ~\cite{Guo}.

Spectra of $\gamma$-rays were detected mainly at scattering angle $\theta$ = 0$^{\circ}$, but registration was done also for scattering angles  $\theta$ = 90$^{\circ}$  (10 resonances), 55$^{\circ}$  and 35$^{\circ}$  (for each angle 4 resonances). The detection was done with a 50 or 75 cm$^{3}$ Ge(Li) detectors incorporated with a CANBERRA preamplifier and a SILENA CICERO 8192-channel analyser, one spectrum was measured with the 145 cm$^{3}$ HPGe detector. The energy resolution of all system was about 1.4 -- 8 keV in the $\gamma$-ray energy region E$_{\gamma}$ = 0.3 -- 12 MeV for all detectors (the energy resolution for the 1333 keV $^{60}$Co peak was about 2.4, 2.2 and 2.1 keV, respectively). Typical examples of high and low energy part of obtained spectra for the 55 and 75 cm$^{3}$ detectors are shown in figure 1 and 2.

\begin{figure}[p]
\caption{Examples of obtained spectra (high energy region)}
\includegraphics*[30mm,0mm][150mm,260mm]{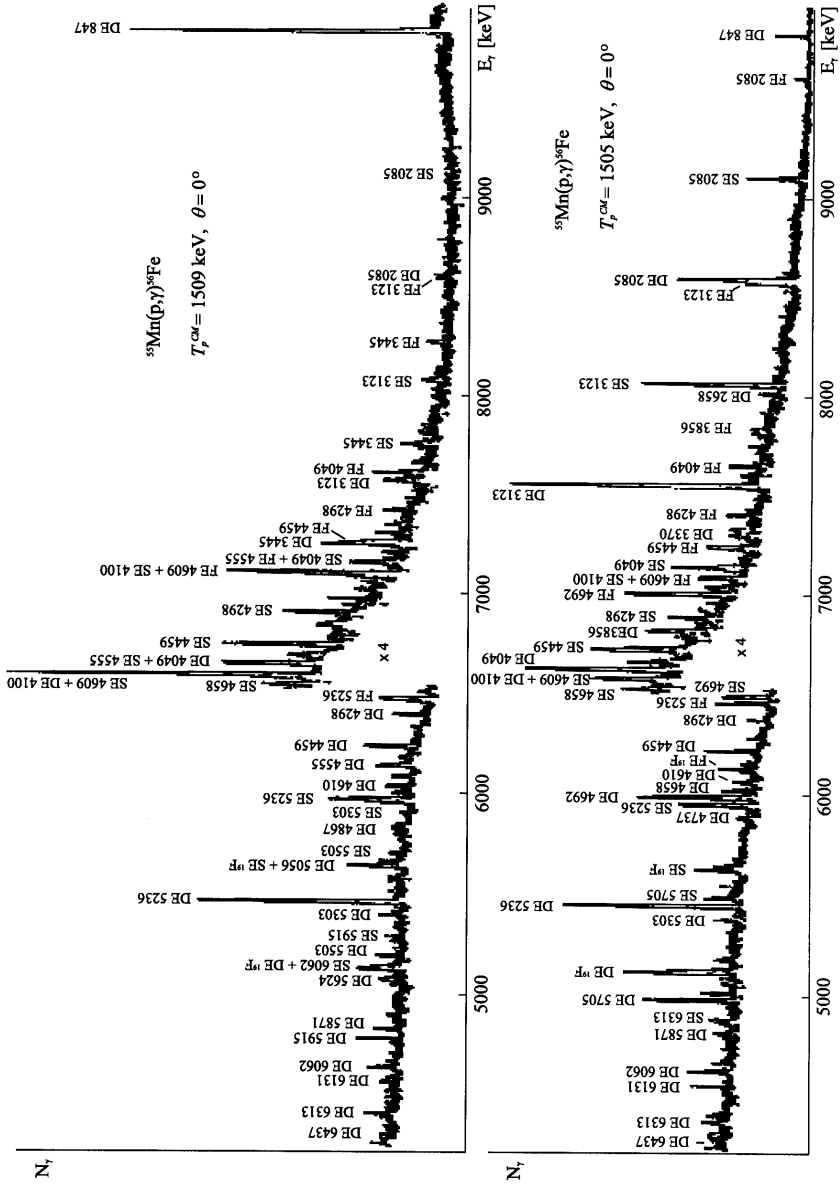}
\end{figure}

\begin{figure}[p]
\caption{Examples of obtained spectra (low energy region)}
\includegraphics*[30mm,0mm][150mm,260mm]{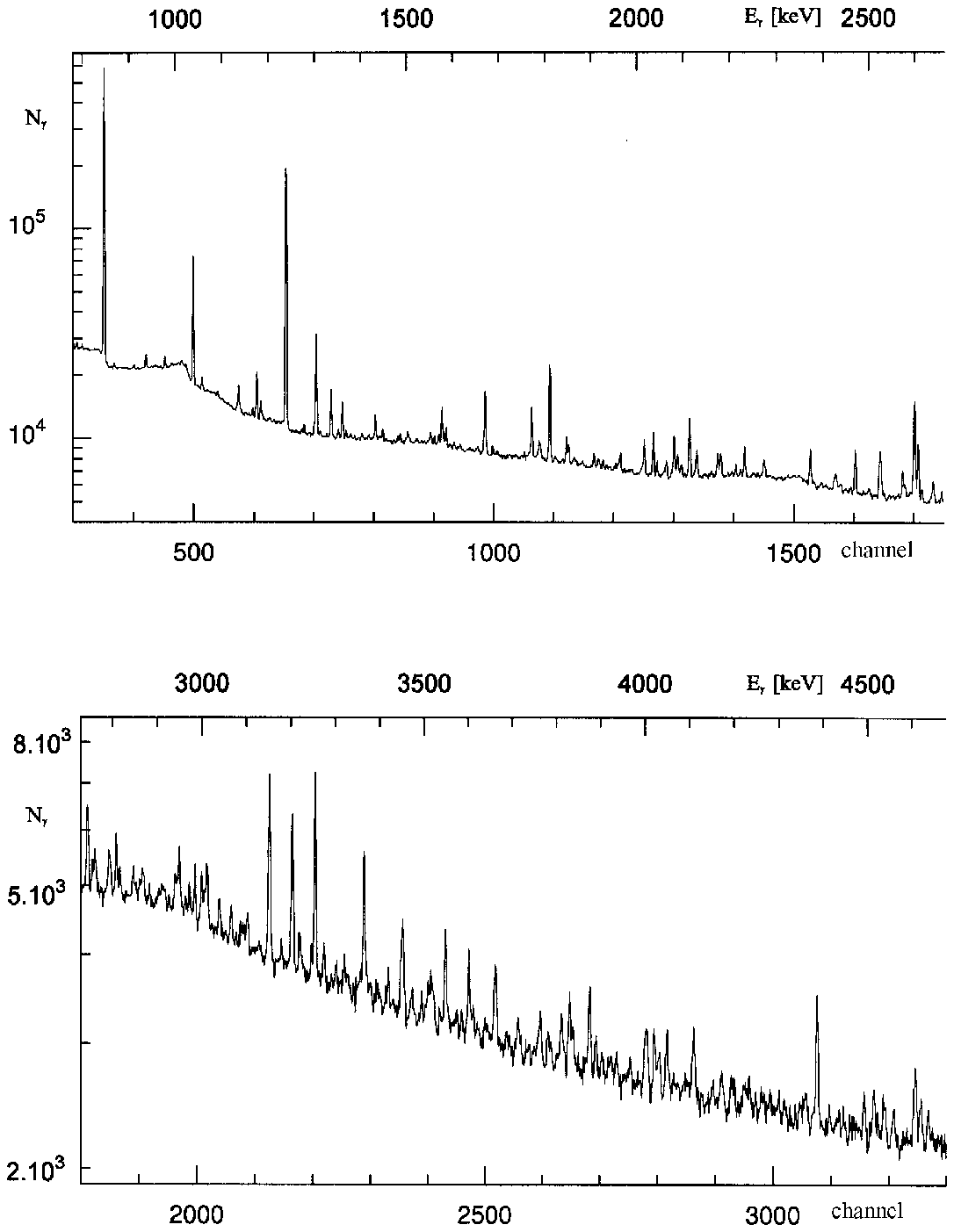}
\end{figure}

Experimental spectra were carefully analysed with the program implemented in the SILENA CICERO analyser and mainly with the special program SPDEMOS from the Nuclear Spectroscopy Department of Institute of Nuclear Physics of Czech Academy of Sciences in Rez ~\cite{SPD}. Big care was paid to the calibration of spectrometric system. External and internal modes were used for both, energy and efficiency calibrations.

\subsection{The energy calibration}

The energy calibration for E$_{\gamma}$ $<$ 3 MeV was done independently by radioactive standards placed in vicinity of the target. For calibration of this region of the spectra were also exploited the 1461 an 2614 keV background transitions in $^{40}$K and $^{228}$Th, respectively, present in all spectra. Known $\gamma$-transitions in the final $^{56}$Fe nucleus and some $\gamma$-rays induced by the protons on target impurities ~\cite{Guo,NDS,Murr}~enabled us to calibrate the system up to E$_{\gamma}$ $\sim$ 6 MeV (6130 keV transition from the $^{19}$F(p,$\gamma$)$^{16}$O reaction). For some expositions  the $\gamma$-ray doublet from the $^{56}$Fe(n,$\gamma$)$^{57}$Fe reaction on surrounding material in laboratory (7631 and 7645 keV)~\cite{NDS2} could be also used, which made it possible to extend direct calibration up to energies about 8~MeV.

Accurate analysis proved that the calibration curve is close to linear one, the deviation from linearity being in all spectrum less or about 1 keV. Nevertheless, all over in accurate determination of the $\gamma$-ray energy the nonlinearity has been considered and the polynomial of third order has been used for the calibration curve.
All $\gamma$-ray energies determined from transitions in the final $^{56}$Fe nucleus have been corrected for the back scattering. For the $\theta$ = 0$^{\circ}$, 35$^{\circ}$ and 55$^{\circ}$ measurements the Doppler effect has been considered and the energy of $\gamma$-ray emitted by flying nucleus has been taken in the known form~\cite{Hami}
\begin{equation}
E_{\gamma} = E_{0} (1 + \frac{v}{c}\cos\theta)
\end{equation}                           
Here {\it c} and {\it v} is the velocity of light and compound nucleus, respectively, $\theta$ is the angle between proton and phonon momentum and {\it E$_{0}$} is the transition energy corrected for the back scattering which equals the $\gamma$-ray energy for {\it v} = 0 or $\theta$~=~90$^{\circ}$.

\subsection{The efficiency calibration}

Intensity I$_{\gamma}$ of single transition has been related to the peak area N, I$_{\gamma}$ = $\varepsilon\cdot$N, where the efficiency $\varepsilon$ of the spectrometer has been taken in the form
\begin{equation}
\varepsilon = exp(a_0 + a_1\cdot\ln{E_{\gamma}} + a_2\cdot\ln{E_{\gamma}} + ...) 
\end{equation}
Parameters {\it a$_{i}$} in (2) have been fitted by the Least Square Method to known experimental intensities. Radioactive standards have again been used in the lowest $\gamma$-ray energy region (E$\gamma$ $<$ 2.5 MeV). Experimental intensities of the $\gamma$-rays at 3395, 4168, 4484 and 6309 keV from the decay of the E$_{p}$ = 1059 keV resonance in the $^{50}$Cr(p,$\gamma$) reaction, measured accurately by Schader et al. ~\cite{Scha}~enabled us to determine the efficiency in the energy region 2.5 -- 6 MeV. Highest part of the spectra can be calibrated only by the primary transitions in the proper $^{55}$Mn(p,$\gamma$) reaction. For determination of their intensity I$_{\gamma}$ we have used the procedure based on the assumption that excitation ({\it s}) and decay ({\it d}) of given {\it k}--th level in final nucleus are in balance. But exploited can be only that levels, for which two basic conditions are fulfilled:
\begin{itemize}
\item The energy of primary to the level is in the region E$_{\gamma}$ $>$ 6 MeV 
\item energy of all other observed transitions to or from the level is in the region calibrated by other manner.
\end{itemize}
The intensity {\it $^{k}$I$_{p}$} of primary transition to which the efficiency have been fitted might then be calculated as
\begin{equation}
^{k}I_{p}=\sum_{j} {^{k}I_{dj}} - \sum_{i\neq p} {^{k}I_{si}}
\end{equation}
where {\it $^{j}$I$_{dj}$} and {\it $^{k}$I$_{si}$} are experimental intensities of secondaries, which deexcite ({\it d}) and excite ({\it s}) the level. However, possible neglection of weak secondaries in (3) results to some uncertainty  of the method.
 
Accuracy of efficiency established for all experimental apparatus is about 20\% for $\gamma$-ray energies E$_{\gamma}$ $>$ 6 MeV and is about 10\% or better for E$_{\gamma}$~$<$~3~MeV.

For higher energies of the $\gamma$-rays creation of the {\it $e^+e^-$} pairs in the detector volume becoms important. Therefore all full energy  (FEP) and escape (SEP and DEP) peaks have been considered in the analysis of the experimental spectra, what also somewhere makes it easier to distinguish different transitions.

\section{Experimental results and used methods}

Experimental $\gamma$-ray spectra of the $^{55}$Mn(p,$\gamma$)$^{56}$Fe reaction are very complex. It is a consequence of high level density in final $^{56}$Fe compound nucleus in the resonant energy region (reaction energy Q = 10184 keV). At the proton beam energy E$_{p}$ $>$ 1 MeV the states with spin J = 0 -- 5 can be excited by the {\it s-}, {\it p-} or {\it d-} capture and the all over average density is expected to be about 5 states per 1 keV ~\cite{Guo}. Therefore, with respect to our proton beam energy resolution (2~--~3~keV), a few levels with different J can participate to one resonance. Nevertheless for many resonances some well defined nuclear states with definite spin and parity might dominate (e. g. isobaric analogue states --- IAS) and determine also the decay of given resonance. Other states, generally with different spin and parity, should also weakly contribute to the resonance and to influence its decay. Important for the resonance structure and decay could be also some special modes of nuclear interaction (e.g. the $\Delta$T~=~1 mixing of the IAS), which result to additional splitting of participating states and again substantially increase a number of bound states achievable from the resonance. All these mixing effects also complicate application of the selection rules for the E1, M1 and E2 electromagnetic transitions, which are expected to be dominant in the $\gamma$-decay.

Spectrum of primaries is overlapped by secondary transitions (``secondaries'') following the decay of bound states of the $^{56}$Fe nucleus. For resulting experimental $\gamma$-ray spectra is then rather a rule that more than one transition participate to given experimental $\gamma$-ray peak, what drasticcaly complicates the analysis of spectra. Therefore, except the energy, also some other indications for identification of individual primaries and secondaries have to be used.

There are no problems in identification of intense primaries to lowest known bound states of the final $^{56}$Fe nucleus (E$_{x}$ $<$ 5 MeV). They may be used for preliminary determination of the energy E$_{r}$ of the resonance from equation
\begin{equation}
E_{r} = E_{\gamma} + E_{x},
\end{equation}
where {\it E$_{x}$} is known energy of the bound state and {\it E$_{\gamma}$} is energy of corresponding primary. From the energy balance (4) might then be identified medium energy primaries to other known bound states. As was already mentioned, individual transitions in the spectra may be sometimes separated by application of all three, full energy an escape peaks. Best tool for identification of individual transitions is, however, increasing of energy E$_{\gamma}$ of primaries with increasing of the proton beam energy E$_{p}$. This energy dependence makes it also possible, in addition to identification of primaries,  to search for new bound states with excitation energy E$_{x}$ assuming that E$_{\gamma}$ in Eq. (4) depends on E$_{p}$. Energy of secondary transitions is independent on E$_{p }$ but their intensity should approximately follow the intensity of primaries, which excite decaying levels. Comparison of both intensities for given level for different resonances is therefore an effective tool for secondaries identification.

To eliminate possible accidental $\gamma$-rays from unknown target impurities we assume that transitions in the $^{56}$Fe nucleus should be observed at more than one (as a rule at five or more) resonances with substantially different energy E$_{r}$. 

As a result of careful analysis we have identified few hundreds of primary and secondary transitions connected with more than two hundreds of bound states in the $^{56}$Fe nucleus.

Application of described criteria made it possible to identify decay of the resonances to bound states up to about 8 MeV of excitation energy E$_{x}$. However, while for E$_{x}$  $<$ 6 MeV we believe that we have found rather all  states, for higher E$_{x}$ only a part of corresponding primaries has been found. For E$_{x}$~$>$~8~MeV the primaries and secondaries are strongly overlapped each other and their separation is practically impossible. Nevertheless, evidently none of the studied resonances decays heavily to any bound state in this energy region.

Observed transition intensity for each measurement has been related to full decay intensity S$_{0}$ of the resonance, which is taken S$_{0}$ = 100\%. Because studied resonances as well as bound states in the $^{56}$Fe nucleus decay very rarely directly to the ground state and practically all decays proceed through the first excited state at 847 keV ~\cite{Ninh}, we have assumed that intensity S$_{0}$ equals the intensity of the 847 keV transition, S$_{0}$ = I$_{\gamma}$(847). Exceptions are the resonances at E$_{p}$~=~1441 and 1454 keV which decay strongly to the ground state ~\cite{Guo}. Here the intensity S$_{0 }$ has been normalized to the sum of intensity of this primary and of the 847 keV transition, $S_{0} = [ I_{\gamma}(847) +  I_{\gamma}(R \longrightarrow g.s.) ]$.

All experimental energies and intensities presented in the work were calculated as weight averages from all observed corresponding values. Experimental errors were calculated independently as the error of average and as a weighted mean square deviation. The higher of both values has been used all over in the paper.

\section{Analysis of the results}

Extensive experimental information concerning the $\gamma$-transition have enabled us to study properties of decaying resonances and of some bound states in the $^{56}$Fe nucleus. Except the excitation energy we have tried to determine for each resonance and for some bound states their decay (branching ratios) and to determine or at least limite their spin and parity. For several bound states also their lifetime has been acquired. Because obtained results are too extensive for one paper, the present paper is devoted only to resonances. Properties of studied bound states should be discussed in following publication.

\subsection{The resonances in the compound $^{56}$Fe system}

Energy E$_{r}$ of resonances has been established for each measurement as weighted average of {\it k} values calculated from (4) for {\it k} levels with well known excitation energy up to E$_{x}$ $<$ 6.5 MeV. Excitation energy E$_{x}$ of the levels in $^{56}$Fe nucleus has been usually taken from literature ~\cite{Guo,NDS}, however also accurate energy of some levels newly observed in our work has been used in this procedure. Energy E$_{\gamma}$ of primaries has been taken from our work and for the $\theta$ = 0$^{\circ}$ measurements has been corected for the Doppler effect (see. Eq. (1)) assuming prompt emission of the photon. Becouse of some uncertainty concerning small nonlinearity of spectrometer the final error of E$_{r}$ for each measurement has been taken as two time the value obtained from experimental $\gamma$-ray spectrum. Final values of E$_{r}$ given in tab. 1 have been then calculated as weighted average from all measurements for given resonance.

Proton beam energy E$_{p}^{CM}$ in the centre-of-mass system corresponding to the resonance energy E$_{r}$ has been calculated from equation
\begin{equation}
E_{p}^{CM} = E_{r} - Q,
\end{equation}
where {\it Q} = (10183.84 $\pm$ 0.17) keV is the reaction energy taken from paper ~\cite{Guo}. Proton beam energy $E_p = E_p^{LS}$ transformed to the laboratory system is in first column of tab. 1 and is used all over for identification of resonances.

Accuracy of calculated energy of resonances has been proved by comparison of our results for the E$_{p}$ = 1441 and 1455 keV resonances with very accurate measurements performed by Guo et al. ~\cite{Guo}. With respect to the experimental errors are the results of both experiments in excellent agreement.

Decay of the resonances is very complex and in average more than 60 primaries to the levels up to excitation energy E$_{x}$ $\sim$ 7 MeV have been observed for each resonance, but somewhere this number exceeds 80. Average radiative strength S$_{\gamma}$ = $\Sigma$(I$_{{\gamma}{i}}$) exhausted by all observed primaries is about 90\% of full radiative strength S$_{0}$, but for some resonances is only about 70\% or less. Therefore with respect to expected error of S$_{\gamma}$ about 10 -- 20\% decay of these resonances to higher bound states (over 7 MeV) exhaust about 10 -- 30\% of full radiative strength S$_{0}$. Nevertheless search for corresponding primaties has not been succesful.

The decay branchings to the levels with excitation energy E$_{x}$ $\leq$ 6.5 MeV acquired for $\theta$ = 0$^{\circ}$ measurements for all resonances are in tab. 2 (except the 1796 keV resonance which was measured only for $\theta$ = 90$^{\circ}$). Because the number of excited states is very high (about 130 levels in this energy region), only that levels for which relative intensity I$_{\gamma}$ exceeds 1\% at least for one resonance are included . The table indicates high number of primaries for each resonance, but in few cases only 3 -- 4 transitions exhaust more than 50\% of full radiative strength. Nevertheless, only for nine resonances have been found strong transitions with relative intensity I$_{\gamma} >$ 10\% and relative intensity for single transition I$_{\gamma} >$ 30\% have been observed only for five resonances.
\begin{table}[p]
{\tiny \begin{tabular}{|r|c|r|r|r|r|r|r|r|r|r|r|}
\hline 
     &E$_{p}$[keV]&1344&1441&1455&1481&1483&1487&1507&1521&1524&1532\\
\hline
E$_{x}$[keV]&$^{d)}$J$^{\pi}$&&&&&&&&&&\\
\hline
\hline
    0&   0$^{+}$ & 0.0&28.5&30.0&    &     &    &    &    &    &     \\      
  847&   2$^{+}$ & 1.6&10.7&13.6& 3.7&  2.1& 9.0& 6.4& 1.4& 1.4&  0.0\\
 2085&   4$^{+}$ & 2.9& 1.5&    & 3.8&  1.8& 6.4& 2.7& 4.0& 1.6&  3.7\\
 2658&   2$^{+}$ & 2.0& 0.7& 1.1& 1.3&  0.3&    & 0.8& 0.8& 0.8&  0.7\\
 2960&   2$^{+}$ & 7.3& 4.2& 2.0& 1.3&  2.5& 4.2&    &    & 1.0&     \\
 3123&   4$^{+}$ & 1.6& 1.8& 0.5& 8.3&  5.1& 7.1&12.0&12.5& 1.4&  7.0\\
 3370&   2$^{+}$ & 8.9& 2.0& 0.6& 0.5&  0.4& 1.7&30.3&    &    &  0.7\\
 3389&   6$^{+}$ &    & 0.6& 0.4&    &  0.3&    & 0.7&    &    &  0.6\\
$^{a}$3445&   3$^{+}$ & 1.4& 0.8& 2.0& 2.9&  0.8& 2.3&    & 3.9& 6.5&  0.5\\
 3600&$^{\alpha}$(1$^{+}$,2$^{+}$)&0.0& 0.8& 1.6& 0.7&  0.4& 0.4& 1.1& 0.0& 0.0&  0.0\\
 3610&  0($^{+}$)&    & 1.2& 3.4& 0.6&     & 0.4&    &    &    &     \\
 3756&   6$^{+}$ &    & 0.4&    &    &     &    & 0.2&    & 0.3&  0.3\\
 3830&   2$^{+}$ & 4.3& 1.0&    &    &     & 1.5& 0.3&    & 0.2&     \\
 3856&   3$^{+}$ & 0.2&    & 0.6& 3.0&  1.7& 1.2& 2.9& 2.1& 2.1&  1.4\\
 4049&   3$^{+}$ & 2.0& 0.2&    & 1.0&  1.0& 2.6&    & 0.8& 0.4&  4.0\\
 4100&   4$^{+}$ & 4.5& 1.3& 1.1& 2.2&  0.5& 0.8& 3.4& 1.1& 0.5&  1.3\\
 4120&   3$^{+}$ & 3.2&    &    & 4.4&  1.3& 0.8& 0.7& 2.7& 1.1&  0.5\\
$^{a}$4298&   4$^{+}$ & 6.5&    &    & 4.0&  7.4& 4.3& 0.8& 1.6& 0.8&  2.6\\
 4395&   3$^{+}$ & 0.9& 0.4&    & 3.2&  1.8& 1.8& 0.6& 0.9& 0.7&  1.0\\
 4401&   2$^{+}$ & 5.1& 3.4&    & 0.5&  0.6&    & 0.6&    &    &     \\
 4458&   4$^{+}$ & 6.4& 0.2&    & 1.9&  3.3& 2.0& 1.5& 2.0& 2.8&  4.8\\
 4509&$^{\beta}$4$^{+}$& 0.6& 1.0& 0.5& 4.0&  3.3& 4.2& 0.5& 1.5& 1.2&  1.4\\
 4540&   2$^{+}$ & 2.2& 4.3& 0.8&    &      &   &    & 0.8& 0.3&     \\
 4555&   4$^{+}$ & 1.2& 0.2& 1.0& 1.0&  0.7& 0.8& 3.4& 1.2& 0.4&  1.2\\
$^{a}$4610&   4$^{+}$ & 3.8& 0.2&    & 0.8&  0.6&    & 0.5& 3.7& 1.6&  2.6\\
 4658&  2-4 & 6.6&    &    & 4.5&  3.8& 2.2& 2.7& 0.7& 1.4&  2.6\\
 4683&4$^{+}$& 8.1& 0.2&    & 0.6&  0.7& 1.8& 0.9& 0.6& 0.9&  1.8\\
 4692&   4$^{+}$ &    & 0.3& 0.4&    &    &  0.9& 1.1& 0.9& 1.3&  8.8\\
 4737&      & 1.2&    & 0.2&    &  0.4&    & 4.3& 4.5& 7.4&  0.9\\
 4867& (1,2)&    &11.4& 8.7&    &  0.4& 1.4&    &    &    &     \\
$^{a}$4878&   2$^{+}$ &    &    &    & 1.9&  2.1& 1.5&    & 0.8& 0.8&     \\
 5023& (1,2)& 3.1& 6.1& 1.9&    &  0.5&    &    &    &    &  0.5\\
 5033&      &    &    &    &    &  0.9& 0.6&    & 1.3&    &  2.5\\
 5038&   4$^{+}$ & 2.1&    & 1.1& 2.6&  1.8& 0.9&    &    &    &     \\
 5056& 4(3)$^{+}$&    &    &    & 1.8&  0.9&    & 2.7&    &    &     \\
 5132&  2-4 & 1.7&    &    & 0.6&  0.6&    & 0.8& 1.3& 1.3&  1.9\\
 5150&      &    &    &    & 1.7&  5.2& 2.2&    & 0.6&    &  0.2\\
 5187&   2$^{+}$ &    & 0.1&    & 0.9&  3.9& 2.3& 0.5& 1.2&    &     \\
 5233& (2,3)& 4.1& 0.5&    & 0.8&  1.1& 0.8& 3.3& 4.2& 1.8&  3.6\\
 5236&  4$^{+}$? &    &    & 0.8& 4.2&  0.6& 0.5&    & 3.7& 4.6&  9.2\\
 5284&      &    &    &    & 1.1&  2.4& 1.4& 0.5&    &    &     \\
 5303&      & 4.4&    &    & 1.6&  0.4&    & 2.0& 1.1& 0.7&  1.8\\
 5452&   4$^{+}$ & 0.5& 0.1& 0.3& 1.9&  1.3& 1.1&    & 1.3& 2.0&  0.6\\
 5488&      & 2.2& 0.2&    & 0.8&  1.0& 0.8&    &    & 1.1&  1.2\\
 5503&      & 0.4& 0.2& 0.5& 1.4&  3.0& 1.9& 1.2& 2.0& 1.4&     \\
 5538&      &    & 1.6&    & 0.5&     &    &    &    &    &     \\
 5562&      & 0.1&    &    & 0.5&     & 1.4&    & 2.4& 1.8&     \\
 5574&   2$^{+}$ & 0.9& 0.4&    &    &     &    &    &    & 0.5&     \\
 5618&      & 1.5& 0.1&    & 1.0&     & 0.6& 1.5& 2.5& 4.5&     \\
 5624&  4-5 & 0.6& 0.1&    & 0.7&  7.7& 4.5& 2.6& 1.7& 1.0&     \\
 5670&      &    & 0.1& 0.3&    &  0.9& 0.8& 0.6& 1.0& 1.1&  1.8\\
 5695&  2-4 & 0.6&    &    &    &     & 0.5& 0.4& 0.4&    &     \\
 5705&      & 1.3& 0.1&    & 1.7&     & 0.7& 0.5& 1.7& 1.5&  5.6\\
 5817&      & 0.1&    &    & 1.4&  2.6& 1.8&    &    &    &  0.6\\
 5863&   4$^{+}$ & 0.5&    &    & 0.8&  0.6& 2.6&    &    &    &     \\
 5871&      & 4.1&    & 0.4& 1.3&  0.8& 2.6& 1.4& 1.2& 1.0&  1.2\\
$^{a}$5915&   2$^{+}$ &    & 0.1&    & 1.8&  5.0& 2.6& 0.7& 1.3& 0.7&  0.8\\
 5934&   2$^{+}$ & 0.2& 0.1& 1.1& 0.5&  0.5&    &    & 0.2& 0.8&  0.4\\
 5987&      &    & 0.5& 0.8& 0.8&  0.8& 0.5&    & 0.7& 0.6&  0.4\\
 6021&      & 1.2&    &    & 0.6&  1.2& 0.4&    &    & 0.6&  1.4\\
 6048&   4$^{+}$ & 0.6&    &    &    &  0.6&    &    & 1.1&    &  0.6\\
 6062&   4$^{+}$ & 2.7& 0.6&    & 1.3&  1.7& 1.7& 1.0& 0.8& 0.9&  3.0\\
 6131&  1-3 & 0.2&    &    &    &  1.8& 0.5& 0.4& 1.2& 0.8&  2.6\\
 6146&      & 0.2&    &    & 0.6&  0.2&    &    & 1.9& 3.4&     \\
 6318&      &    &    &    & 0.9&  0.7&    &    & 1.1& 1.4&  1.3\\
 6387&      &    &    & 0.4& 0.9&     &    & 0.8& 1.4& 2.2&  1.0\\
 6437&      & 0.3&    &    & 0.7&  2.6& 1.5&    & 0.4& 1.4&  1.0\\
\hline
\end{tabular}\\
\vspace{-2mm}
$^{a)}$For some resonances participate close levels with different assignment\\
\vspace{-2mm}
$^{b)}$Measurements with very low statistic, many primaries might be omitted\\
\vspace{-2mm}
$^{c)}$Measurements only for $\theta$ = 90$^{\circ}$\\
\vspace{-2mm}
$^{d)}$Assignment J$^{\pi}$ is taken from ~\cite{Guo},~\cite{NDS}~and from our analysis of secundary decay of levels\\
\vspace{-2mm}
$^{\alpha}$From our analysis J$^{\pi}$ = 2$^{+}$\\
\vspace{-2mm}
$^{\beta}$Assignment from our analysis, in ~\cite{NDS}~J$^{\pi}$ = 3$^{-}$\\
}
\caption{The decay branching of the resonances in $^{56}$Fe}
\end{table}

\addtocounter{table}{-1}

\begin{table}[p]
{\tiny \begin{tabular}{|r|c|r|r|r|r|r|r|r|r|r|r|r|}
\hline 
&E$_{p}$[keV]&1536&1579&1687&$^{b}$1696&1727&1734&$^{b}$1761&1773&$^{c}$1796&$^{c}$1801&$^{c}$1807\\
\hline
E$_{x}$[keV]&$^{d)}$J$^{\pi}$&&&&&&&&&&&\\                                                
\hline
\hline       
   0& 0$^{+}$&  0.2& &&&&&&&&&\\                                                
 847& 2$^{+}$& 13.2& 4.6& 7.6& 3.0&1.1 &0.8& 1.6&3.0& 1.1& 1.4& 37.1\\
2085& 4$^{+}$&  0.8& 3.5& 8.1& 2.0&2.4& 9.3& 6.5&10.3&1.7& 9.2&  6.5\\
2658& 2$^{+}$&  0.4& 2.5& 2.7& 1.2&0.4&    &    &4.4&    & 0.3&  4.8\\
2960& 2$^{+}$&     & 1.7& 6.3& 4.2&0.5& 1.4&    &1.0&    & 0.3 &    \\
3123& 4$^{+}$&     & 3.8& 6.0& 8.0&2.0& 1.5& 6.9&2.0& 1.3& 1.6&  6.7\\
3370& 2$^{+}$&     & 3.0& 0.3 &1.8&1.1&    & 2.6&1.7&    & 0.5&  1.7\\
3389& 6$^{+}$&  0.6& 1.4&     &   &3.7& 2.9&    &   &    & 1.9&  0.7\\
$^{a}$3445&3$^{+}$&  2.1& 4.6& 4.6& 1.3&0.5 &2.9& 1.5&3.6& 0.8& 1.4&  2.0\\
3600&$^{\alpha}$(1$^{+}$,2$^{+}$)&0.6&5.6& 4.7& 1.4&    &   &    &0.8 &   &    &    \\
3610& 0($^{+}$)&0.2&    &    &    &    &   &    &    &   &    &     \\
3756& 6$^{+}$&     & 2.0& 0.8&    &5.8& 1.0& 1.9&    &   & 0.8&  1.0\\
3830& 2$^{+}$&  0.1& 0.8&    &    &0.4 &   & 2.8&    &   &    &     \\
3856& 3$^{+}$&  0.6& 1.2& 1.4&    &0.5 &   & 2.5&1.9 &0.7& 0.5&     \\
4049& 3$^{+}$&  1.7& 4.5& 4.1& 1.5&0.4& 1.2& 2.2&2.2 &   & 4.7&  3.5\\
4100& 4$^{+}$&  9.0& 1.3& 1.7& 0.9&1.4& 2.7& 2.9&2.7& 2.1& 5.4&  3.1\\
4120& 3$^{+}$&  1.2& 6.1& 7.7&    &0.9 &1.1&    &1.1 &    &0.6 & 1.7\\
$^{a}$4298&4$^{+}$&  3.1& 0.6& 5.2& 2.2&3.0  &  &    &5.5 &2.6 &4.2&  1.4\\
4395& 3$^{+}$&  0.8& 0.7& 1.1& 2.7&1.8& 0.9&    &1.9 &    &   &  0.3\\
4401& 2$^{+}$&  0.2& 1.8& 3.8&    &   &    &    &    &    &   &  0.3\\
4458& 4$^{+}$&  5.6& 4.3& 3.4& 1.2&3.0& 2.5&    &1.6& 0.6& 0.9&  1.2\\
4509&$^{\beta}$4$^{+}$&1.6&1.9& 0.6& 0.8&0.5&    &    &5.5 &4.0& 2.4&  0.9\\
4540& 2$^{+}$&     &    & 0.7& 2.6 &  &    &    &    &0.2 &0.6 &    \\
4555& 4$^{+}$&  3.0& 2.1& 0.7&    &4.4 &0.6 &   &   &10.2&10.8&  3.7\\
$^{a}$4610&4$^{+}$&  2.9& 1.3& 1.7& 1.9&0.8 &2.5&    &0.9& 4.9& 6.9&  1.8\\
4658&2-4&  2.4& 2.7& 2.7&    &    &   &    &1.5& 1.6& 2.4&  0.7\\
4683&4$^{+}$&1.6&3.5& 1.8& 1.4&1.3& 2.2& 2.3&1.9& 1.1& 3.3&  3.2\\
4692& 4$^{+}$&  2.1&    & 1.9& 2.1&2.1& 4.4& 2.1&1.7& 0.7& 2.0&  0.7\\
4737&   &  2.2& 4.8& 2.4&    &    &   &     &  & 1.2& 1.2 & 1.7\\
4867&(1,2)&2.0& 1.5& 4.1& 4.7&    &   &    &1.1 &   &     & 2.0\\
$^{a}$4878&2$^{+}$&  1.0& 1.1& 1.6& 0.7&2.2 &   &    &0.7 &0.4 &0.6&  5.3\\
5023&(1,2)&0.7& 1.7& 1.2&    &    &   & 2.4 &   &    &   &     \\
5033&   &  1.7& 0.2&    & 2.4&1.0& 3.3& 2.4&    &    &    &    \\
5038& 4$^{+}$&  0.9& 1.6& 1.2&    &0.4 &   & 3.1&    &0.9 &0.9 &    \\
5056&4(3)$^{+}$&2.2& 0.6& 0.8&    &1.2 &2.3 &    &   &    &    &    \\
5132&2-4&  1.1& 15.5&11.5&0.8&2.6& 1.3&    &2.2& 1.4& 3.0&  3.5\\
5150&    &  0.3& 1.1& 0.3&    &0.9&    &    &0.8 &   & 0.7 & 0.3\\
5187& 2$^{+}$&  0.7& 0.1&    &    &0.3 &3.3 &   &0.8& 0.4 &    &    \\
5233&(2,3)&8.8& 0.4&    & 2.2&1.0& 1.8&    &5.3& 0.4& 0.9&  8.7\\
5236&4$^{+}$?&  9.3& 1.2& 1.2&    &   &    & 1.3&    &0.4 &0.8&  1.1\\
5284&   &  0.4& 0.4 &   &    &2.5 &0.7 &   &    &    &1.1 &    \\
5303&   &  1.5& 4.0& 2.1& 1.5&    &1.4 &1.4&3.6& 2.6& 4.1&  1.0\\
5452& 4$^{+}$&  0.6& 0.2 &0.1&    &1.5&    &    &1.0& 0.2& 0.7&     \\
5488&   &     & 1.8 &1.8 &0.6&2.3&    &    &1.2 &1.4& 1.1&     \\
5503&   &  2.2& 1.2&&        &0.7& 0.6&    &    &0.7& 1.5 & 3.1\\
5538&   &  0.1&    &     &   &   &    &    &    &   & 0.7 &    \\
5562&   &  0.8& 0.3& 0.2 &   &   &    &    &    &   & 2.0 & 0.2\\
5574& 2$^{+}$&     & 2.1 &0.3&    &   &    &    &    &   &     &    \\
5618&   &  0.8& 1.1 &0.2 &   &   & 1.6  &  &0.7& 0.8& 0.9&  0.7\\
5624&4-5&  1.6& 1.3& 0.4&    &   &    &    &0.4 &   & 0.3 & 0.7\\
5670&   &  0.5& 0.9& 0.8&    &   &    &    &    &   & 0.9 &    \\
5695&2-4&  0.2& 0.9 &2.4 &0.7 &  & 0.9&    &0.6&    & 1.1 &    \\
5705&   &     &    & 1.5 &   &1.6 &2.5 &   &1.6 &0.5& 1.0&  4.9\\
5817&   &     & 0.6&     &   &    &    &1.4 &   &   & 0.6 &    \\
5863& 4$^{+}$&     & 0.3 &0.9 &0.8&1.5&     &    &   &   &     &    \\
5871&   &  1.5& 2.2 &2.8 &1.9&0.5& 2.1& 1.7&1.3 &   & 0.6&  0.2\\
$^{a}$5915&2$^{+}$&  2.3& 0.7& 1.2&    &   &    &    &0.7&    & 2.6 & 1.7\\
5934& 2$^{+}$&  0.3& 0.9& 1.0& 1.6 &   &1.7&    &    &   &     &    \\
5987&   &     & 1.2& 0.7&    &0.5 &   &    &    &   &     &    \\
6021&   &  0.1&    & 0.1 &0.5&0.4&    &    &    &   &     &    \\
6048& 4$^{+}$&  0.6& 1.0& 1.6& 0.7&   &    &    &    &    &    &    \\
6062& 4$^{+}$&  1.8& 1.4 &1.0 &   &0.6&    &    &1.1 &    &1.1 & 1.5\\
6131&1-3&  0.8& 0.2& 1.4&    &1.6&    &    &    &0.4 &0.2&     \\
6146&   &  0.4& 0.3& 0.9&    &   & 1.4 &   &    &0.6& 0.8&     \\
6318&   &  0.2&    &    &    &   &     &   &1.0 &1.3& 1.5&     \\
6387&   &     &    &    & 1.2&0.4&     &   &    &   &    &     \\
6437&   &  0.8& 0.6 &1.2& 1.2&0.7& 1.0 &1.2&    &   & 0.8&     \\
\hline
\end{tabular}\\
\vspace{-2mm}
$^{a)}$For some resonances participate close levels with different assignment\\
\vspace{-2mm}
$^{b)}$Measurements with very low statistic, many primaries might be omitted\\
\vspace{-2mm}
$^{c)}$Measurements only for $\theta$ = 90$^{\circ}$\\
\vspace{-2mm}
$^{d)}$Assignment J$^{\pi}$ is taken from ~\cite{Guo},~\cite{NDS}~and from our analysis of secundary decay of levels\\
\vspace{-2mm}
$^{\alpha}$From our analysis J$^{\pi}$ = 2$^{+}$\\
\vspace{-2mm}
$^{\beta}$Assignment from our analysis, in ~\cite{NDS}~J$^{\pi}$ = 3$^{-}$\\
}
\caption{The decay branching of the resonances in $^{56}$Fe {\it (continued)}}
\end{table}

As was already mentioned in part 3, the resonances with spin J = 0 -- 5 in the $^{56}$Fe compound nucleus should be excited by the $^{55}$Mn(p,$\gamma$)$^{56}$Fe reaction. Some of these resonances are the Isobaric Analogues (IAS) of the lowest states in the $^{56}$Mn nucleus \cite{NDS}. Results of former attempts to identify these resonances have been collected in Nuclear Date Sheets ~\cite{NDS}~and are also included in tab. 1.

Former assignment of the IAS at proton beam energy E$_{p}$ = 1344, 1441, 1455 and 1532 keV with spin and parity J$^{\pi}$ = 3$^{+}$, 1$^{+}$, 1$^{+}$ and 4$^{+}$, respectively seems to be unquestionable and the 3$^{+}$ and 4$^{+}$ assignment is also in agreement with our measurement of angular distribution of primary $\gamma$-rays ~\cite{Trun}. Therefore we have considered these values as definite all over in present work. For all other studied resonances we have tried, beside the energy, to determine also spin--parity assignment J$^{\pi}$. For this purpose we have exploited primary $\gamma$-ray spectra and applied the E1, M1 and E2 selection rules for transition to lower levels. However, as a main tool for determination of J-value we have used two testing methods.
     
{\it $<$J$>$ METHOD.} There was shown in papers \cite{Book,Kiks}, that spin of the resonance should be an average $<$J$>$ of spins of the states, to which the resonance decays. The method described in ~\cite{Kiks}, which takes into account the number of levels with given spin, has been therefore applied in present work for all studied resonances. Carried analysis leads to expected spin for the 1344, 1441 and 1532 keV resonances while J = 2 has been obtained for the 1455 keV one. Because the value J = 1 for last resonance is rather sure, the discrepancy might be probably connected with admixture of the states with different spins in the compound state and also might show the limit of the $<$J$>$ method.

{\it PC(J1,J2) METHOD.} As the probability of $\gamma$-transition is very sensitive to the transition multipolarity, radiative properties of different resonances with the same spin J can be expected to be rather similar. Cameron~\cite{Came} exploited this similarity for destination of unknown spin of resonances in the proton capture reactions by inclusion for comparison of decay of two resonances with spins J$_{1}$ and J$_{2}$ the ``parentage coefficients'' PC(J$_{1}$,J$_{2}$)
\begin{equation}
PC(J_{1},J_{2}) = \sum_i {^1a_i}.{^2a_i}.
\end{equation}
Here coefficients {\it $^{k}$a$_{i}$} are related to relative intensities {\it $^{k}$I$_{{\gamma}{i}}$} of transitions to {\it i}-th\\ bound states and are normalised to 1. It was shown in ~\cite{Came}, that the value of coefficient PC(J$_{1}$,J$_{2}$) decreases with increasing difference between J$_{1}$ and J$_{2}$ and is close to 1 if both values J$_{1}$ and J$_{2}$ are the same. We have applied the method for comparison of radiative properties of studied resonances with properties of the J$^{\pi}$ = 3$^{+}$, 1$^{+}$ 1$^{+}$ and 4$^{+}$  resonances at E$_{p}$ = 1344, 1441, 1455 and 1532 keV, respectively. The coefficients {\it $^{k}$a$_{i}$}  in (7) for {\it k}-th resonance have been calculated in the form 
\begin{equation}
^ka_i=\sqrt{\frac{^kI_{\gamma i}}{E^n}}
\end{equation}
where {\it $^{k}$I$_{{\gamma}{i}}$} is the intensity of {\it i}-th transition and power n = 0 -- 5 has been tested. Application to known resonances indicates that except for n = 5 all results of calculations are close one to other and show that coefficients PC(J$_{1}$,J$_{2}$) are about 0.8 for difference $\Delta$J = 0, about 0.4 -- 0.6  for $\Delta$J = 1 and 2 and are $\leq$~0.3 for $\Delta$J $\geq$ 3. This values of coefficients indicate that only spin J $\geq$ 4 can be determined by comparison of examined resonances with the J$^{\pi}$ = 1$^{+}$ ones while spin J = 2 and 3 of the resonances can hardly be distinguished.

Both, $<$J$>$ and PC(J$_{1}$,J$_{2}$), methods might be influenced by admixture of the states with different J in individual resonances and therefore their application might be in some extent limited. Nevertheless, the methods offer the basic information for spin assignment of the resonances.
 
Accepted assignment of resonances given in sixth column of table 1 is based, except on both methods, also on full character of decay of resonances and is also supported by analysis of decay of resonances to individual bound states. Identification of IAS is also tested by comparison of difference between energy of considered IAS and the J$^{\pi}$ = 3$^{+}$ IAS at E$_{p}$ = 1344 keV with the excitation energy of parentage state in the $^{56}$Mn nucleus.

\section{Discussion}

\subsection{Properties of the resonances}

Accurate determination of resonances energy E$_{r}$ has pointed out that five of the main resonances at proton beam energy E$_{p}$ = 1344, 1536, 1687, 1727 and 1801 keV, which are strongly expressed in the (p,$\gamma$) excitation function ~\cite{Ster}~are probably very close doublets. Although separation energy does not exceed 1~keV, the difference is always 4 -- 5 times the combined error and therefore the separation seems to be real regardless of very similar decay of both components. Reality of splitting has been supported by other resonances for which 2 -- 6 independent measurements have been done even for different scattering angles. Here, in contradiction with quoted resonances, all energies calculated from individual measurements are in limit of experimental errors equal. The energies of splitted resonances are included as a note in to the table 1 but in the discussion of other properties of the resonances the splitting has not been considered.

Attribution of the spin--parity assignment to individual resonances included in seventh column of Tab. 1 has been based on the methods described in part 4.1. Because the properties of some resonances are raher close, discussion of the results of analysis has been done separately for several groups of the resonances. Summary of all analyzed resonances is presented in Tab. 1 on page 2.

\subsubsection{The standard 1344, 1441, 1455 and 1532 keV resonances.}

As it was alredy shown, we have used these resonances, assigned in papers ~\cite{Saka,Peel,Mehn,Ahme} as the 3$^{+}$, 1$^{+}$, 1$^{+}$ and 4$^{+}$ IAS in the $^{56}$Fe nucleus, as standard for comparison with other measured resonances and for proving the $<$J$>$ and PC(J$_{1}$,J$_{2}$) methods (see part 4.1). Global properties of spectra of primary $\gamma$-transitions exciting the levels up to E$_{x}$ $\sim$ 7 MeV have been also used as standard, which are expressively different for individual resonances.
\vspace{-1mm}
\begin{itemize}
\item[-]{While for the 1$^{+}$ and 3$^{+}$ resonances observed strength S$_{\gamma}$ is close to 100\% of full radiative strength S$_{0}$, it is only about 80\% for the 4$^{+}$ resonance.}
\item[-]{Spectra of the $1^{+}$ resonances are rather needy, but more then 60\% of the radiative strength is exhausted by only four transitions (to the ground state, 847, 4867 and 5023 keV levels). Fully different are both other spectra. Here the observed radiative strength S$_{\gamma}$ of primaries to the excitation energy region E$_{x}$ $<$ 7 MeV is distributed between more then 80 transitions but intensity of none of them exceeds 10\%.}
\end{itemize}
\vspace{-1mm}

Observed differences in the primary spectra are evidentlly a result of properties of resonant and low-lying states in the $^{56}$Fe nucleus. Therefore assuming that character of the primary spectra is determined by resonance's J$^{\pi}$ charakterisitics, global properties of individual spectra might be in some extent exploited for discussion of the assignment of corresponding resonances.

\subsubsection{The high spin resonances.}

Comparison with the J$^{\pi}$ = 1$^{+}$ resonances at 1441 and 1455 keV by the PC - test has pointed out that spin of the resonances at proton beam energy E$_{p}$~=~1521, 1524, 1727, 1734, 1761, 1796 and 1801 keV should be J $\geq$ 4. Except the 1524 keV resonance ($<$J$>$ = 3, 4) this J-value follows also from the $<$J$>$-test. High spin of these resonances is also in agreement with character of the spectra of primaries. All resonances decay strongly to the J$^{\pi}$ = 3$^{+}$ and 4$^{+}$ levels while no transitions to the J$^{\pi}$ = 0$^{+}$ and 1$^{+}$ states have been found. Observed radiative strength S$_{\gamma}$ of all these resonances is low and is only S$_{\gamma}$ $\sim$ 70\% of the full strangth S$_{0}$ or less. Therefore at least about 30\% of the full radiative strength S$_{0}$ should excite the levels in higher region of the excitation energy, but corresponding primaries have not been found. Irrespective to these common properties the decay of individual resonances differs as will be discused later.

We have not found any former publication concerning the spin-parity assignment of any of these resonances. 

{\underline {\it The 1521 and 1524 resonances.}} From table 2 one can see that both energetically clearly separated resonances have a slightly different decay, however, main properties of them are common. The strongest transitions are that to the 2085, 3123 and 4610 keV J$^{\pi}$~=~4$^{+}$ levels, but of comparable intensity are transitions to the 3445 and 3856 keV 3$^{+}$ states. Decay to the 2$^{+}$ levels is generally rather weak or missing at all. As both resonances decay also to the 3756 keV 6$^{+}$ level and to the level at 5122 keV, assigned in ~\cite{NDS}~ as J$^{\pi}$ = 5$^{-}$ their assignment as J$^{\pi}$ = 4$^{+}$ ones is rather sure. The resonances might be, together with the 1532 keV one (part. 5.1), split isobaric analogue of the 212 keV 4$^{+}$ state in the $^{56}$Mn nucleus, what is supported by very similar decay of all three resonances (the mutual parentage coefficients PC(J$_{1}$,J$_{2}$) are high and are about 0.8 for all cocorresponding measurements).

{\underline {\it The 1727 and 1734 keV resonances.}} The resonances correspond to the 1728 keV resonance observed in the $^{55}$Mn(p,$\gamma$) excitation function, which is probably a doublet ~\cite{Ster}. The splitting was observed in the (p,n$_{1}$$\gamma$) channel while in the inelastic (p,p$_{1}$$\gamma$) channel the resonances were not surely identified~\cite{Ster}. The splitting indubitably follows from the radiative decay studied in present work (7 independent measurements at E$_{p}$ = 1727 and 1733 keV.

The $<$J$>$ and PC criteria indicate for both resonances spin J $\geq$ 4 in agreement with their strong decay to some 4$^{+}$ and rather strong decay to both known 6$^{+}$ levels. However, transition to the 5122 keV 5$^{-}$ level has not been found. Except the decay of the 1734 keV resonance to the 3445 keV 3$^{+}$ state few observed transitions to the 2$^{+}$ and 3$^{+}$ states are rather weak. No transition to the 0$^{+}$, 1$^{+}$ and 3$^{-}$ levels has been observed. 

In the parent $^{56}$Mn nucleus is known the J$^{\pi}$ = 5$^{+}$ state at 335.5 keV of excitation energy ~\cite{NDS}. Proton beam energy  E$_{p}$ of corresponding analogue resonance in the $^{56}$Fe nucleus should be about 1690 keV. Because spin of other resonances observed in this energy region is indubitaly J $<$ 4 and we have not found any other candidate for the 5$^{+}$ IAR, we have tentatively ascribed  the 1727 and 1734 keV resonances as the split T = 3 analogue of the 355.5 keV 5$^{+}$ state in $^{56}$Mn. But no observed transition to the 5122 keV 5$^{-}$ state, considerable intensity of decay to some 3$^{+}$ states as well as not observed resonance in the inelastic scattering (p,p$_{1}$$\gamma$) channel make this assignment rather doubtful.

{\underline {\it The 1801 keV resonance.}} Spin of the resonance J $\geq$ 4 follows from both, the $<$J$>$ and PC tests and is supported by observed transitions to both known 6$^{+}$ states at 3389 and 3756 keV. Only few rather weak transitions to the 2$^{+}$ levels have been found. High spin of the resonance is also supported by rather low value of observed radiative strength S$_{\gamma}$, which is about 70\% of full radiative strength S$_{0}$. The splitting of the resonance noted in tab. 1 should be considered as proved because the energy difference, calculated from six independent measurements, exceeds about 4 times the combined error. But in frame of experimental errors the decay of both members of the doublet is identical.

{\underline {\it The 1796 keV resonance.}} The resonance is the last one in the group of the resonances for which both, the $<$J$>$ an PC tests, indicate high spin, J $\geq$ 4, however the decay of resonance with expressively dominant transition to the 4555 keV 4$^{+}$ state, is rather strange. Extremally low is the radiative strength S$_{\gamma}$, which exhausts only about 50\% of full radiative strength S$_{0}$. Except one to unassigned 5303 keV level all other stronger transitions excite the J$^{\pi}$ = 4$^{+}$ states while none of both transitions to the 6$^{+}$ states has been found and very scare or fully missing are primaries to the 2$^{+}$ states. Absence of trasitions to the 6$^{+}$ and 2$^{+}$ states could indicate odd parity of the resonance, but weaknes of decay to known 3$^{+}$ states is in some extent in contradiction with such assignment. Nevertheless, the J$^{\pi}$ = 4$^{-}$ assignment is included in table 1 as a most probable.

{\underline {\it The 1507 keV resonance.}} Last resonance which might be included into this chapter is the 1507 keV one, although the PC-test is less clear. Resonance decays very intensely to some 2$^{+}$ and 4$^{+}$ levels, the intensity of primary to the 3370 keV 2$^{+}$ level being as high as I$_{\gamma}$ $>$ 30\%. Rather intense is also decay to some J$^{\pi}$ = 3$^{+}$ levels and also weak decay to both 6$^{+}$ levels has been found. Observed radiative strength S$_{\gamma}$ is high and is close to 100\% of full radiative strength S$_{0}$, similarly as it is for the E$_{p}$ = 1344 keV 3$^{+}$resonance. Both, the 2$^{+}$ intensity and S$_{\gamma}$ value indicate that spin of the 1507 keV resonance is J = 3 rather than J = 4. Transitions to the 6$^{+}$ levels could then be connected with the high spin admixture in the resonance. The extremely intense transitions to some 2$^{+}$ and 4$^{+}$ states may be explained, at least partly, if they are of the E1 character, but than parity of the resonance must be negative. Negative parity might be also a reason that the PC(1507,3$^{+}$) and PC(1507,4$^{+}$) coefficients are both about 0.5~\cite{Came}~. Therefore we considere the J$^{\pi}$~=~3$^{-}$ assignment of the 1507 keV resonance as rather sure (see.  tab. 1).

{\underline {\it The 1761 keV resonance.}} The resonance has been measured with a very bad statistics and many weaker primaries are evidently omitted. Therefore observed radiative strength S$_{\gamma}$ is only about 50\% of full radiative strength S$_{0}$. The $<$J$>$-test indicates J = 4 value for the spin of the resonance, in rough agreement with rather uncertain PC-test pointing J = 3 or 4. As a strongly dominating primary transitions (I$_{\gamma}$ $\sim$ 6 -- 7\% ) excite 2085 and 3123 keV 4$^{+}$ levels in $^{56}$Fe nucleus and also decay to the 3756 keV 6$^{+}$ state is clearly observed, we tentatively assume that the assignment of the resonance is J$^{\pi}$ = 4$^{+}$, as is included in table 1.\\

\subsubsection{The other resonances}

The values of all PC(J$_{1}$,J$_{2}$) coefficients calculated for the resonances at E$_{p}$~=~1481, 1483, 1487, 1536, 1679, 1687, 1696, 1773 and 1807 keV and for the J$_{2}^{\pi}$ = 1$^{+}$ reference resonances are PC $\sim$ 0.40 what indicates $\Delta$J = 2 change of the spin. PC coefficients related to the 1321 keV 3$^{+}$ and 1504 keV 4$^{+}$ resonances indicate preferably the J~=~3 value of the spin but often the J~=~4 should not be exluded. Similar results follow also from the $<J>$ test, which for the 1679 keV resonance even preferes J = 4 value. Except the 1774 and 1807 keV resonances all other ones are characterised by observed radiative strength S$_{\gamma}$~$\sim$~100\% of full S$_{0}$ value and, except the 1807 keV one, by rather intense decay (I$_{\gamma}$ $>$ 1\%) to many bound states. All these common properties resemble the decay properties of the J$^{\pi}$~=~3$^{+}$ 1344 keV resonance, nevertheless, decay of individual resonances is rather different and need some further discussion.

{\underline {\it The 1481, 1483 and 1487 keV resonances.}} The mutual PC coefficients of all three resonances are PC $\sim$ 0.85. PC coefficients related to the 4$^{+}$ and 3$^{+}$ known resonances are about the some values, PC $\sim$ 0.65 -- 0.70. Decay of individual resonances is somewhat different, nevertheless all most intense primaries excite some 4$^{+}$ levels and for the 1487 keV resonance also the 847 keV 2$^{+}$ level. Primaries to other 2$^{+}$ and to the 3$^{+}$ levels are slightly weaker. Except one very weak transition to the 3389 keV level no other primaries to the 6$^{+}$ levels have been found. All these information are in rough agreement with spin J = 3 for all three resonances but parity becomes uncertain.Nevertheless, as was noted in~\cite{Came}, the PC coefficients might be affected also by parity of the resonance. Therefore we assume that rather small difference of the PC(X,3$^{+}$) and PC(X,4$^{+}$) coefficients could point that parity of all three resonances is odd and J$^{\pi}$ = 3$^{-}$ assignment is in tab. 1.

{\underline {\it The 1536 keV resonance.}} In all three, (p,p$_{1}$$\gamma$), (p,n$_{1}$$\gamma$) and (p,$\gamma$) channels is observed very strong resonance at E$_{p}$ = 1536 keV, which is in the (p,$\gamma$) channel more than 2-times stronger than any other resonance in the proton beam energy region E$_{p}$ $\sim$ 1.2 -- 1.9 MeV. Radiative decay of this resonance was studied in former papers ~\cite{Frid,Kaz1,Peel}~and our measurements are in rough agreement with them. Peel et al. ~\cite{Peel}~assigned the resonance as the J$^{\pi}$ = 2$^{+}$ isobaric analogue of the 215 keV state in the $^{56}$Mn nucleus. Weak transition to the ground state of the $^{56}$Fe nucleus observed in all our measurements as well as the strongest decay to the 847 keV 2$^{+}$ first excited state (I$_{\gamma}$ $\sim$ 13\%) support this assignment, however all other our information are in contradiction with it. Both, the $<$J$>$ and PC tests, indicate J = 3 or even J = 4 value. More, this spin value follows from analysis of angular distributions for two cascades and for six primaries, measured in our group ~\cite{Trun}. Because also weak transition to the 3389 keV 6$^{+}$ level has been undoubtfully observed, we have concluded that the resonance is a composition of the J$^{\pi}$ = 2$^{+}$ isobaric analogue of the 215~keV level in the $^{56}$Mn nucleus and of other dominating state for which the J$^{\pi}$ = 4$^{+}$ assignment seems as a most probable. Such assignment is also included in table 1.

Observed intensity S$_{\gamma}$ of primary decay of the resonance exhausts about 90\% of full radiative strength S$_{0}$. But, although the decay of the resonance is very complex (I$_{\gamma}$ $>$ 1\% for about 30 primaries is ), more than 45\% of total radiative strength S$_{0}$ is exhausted by only 5 strongest transitions to the 2$^{+}$ and 4$^{+}$ levels.

As is noted in tab. 1, one measurement from 6 ones performed for the resonance, indicates splitting of the resonance. Energy difference exceeds more than four times the combined experimental error what makes the splitting rather proved. Nevertheless, the decay spectra are exactly the same for all measurements (the cross-over PC coefficients for all measurements are PC $\geq$ 0.9), what indicates that both components might be a fine structure of one resonance.

{\underline {\it The 1679, 1687 and 1696 keV resonances.}} First two resonances were interpreted by Chaturvedi et al. ~\cite{Chat}~as the fine structure of the 4$^{+}$ isobaric analogue of the 335.5 keV while the 1696 keV resonance as the 3$^{+}$ analogue of the 341 keV states in the $^{56}$Mn nucleus, respectively. But the 4$^{+}$ assignment is in contradiction with later publication ~\cite{NDS}~where the 335.5 keV level in $^{56}$Mn is assigned as a pure 5$^{+}$ state. Angular distributions for the 1679 keV resonance measured in our group (two cascades and six primaries) are consistent only with spin J = 3 while the J = 4 value is excluded.

Radiative decay of the 1696 keV resonance measured in our group is in rough agreement with former publications ~\cite{Frid,Chat}. But our measurement was done with very low statistics an weaker transitions might be omitted. This is probably a reason of low observed radiative strength (S$_{\gamma}$ $\sim$ 70\% of S$_{0}$) and very uncertain PC test for this resonance. Nevertheless, none of our information are in strong contradiction with the J$^{\pi}$ = 3$^{+}$ assignment of the resonance done in ~\cite{Chat}.

Radiative decay of both other resonances measured by us is in excellent agreement with ref.~\cite{Frid}~and in rough agreement with ref.~\cite{Chat}. Both primary spectra are very similar (the mutual PC coefficients are PC $>$ 0.85) and most complex from all resonances measured in present work (for 35 primaries I$_{\gamma}$ $>$ 1\%). Both resonances decay very intensely to the 5132 keV level with uncertain assignment (I$_{\gamma}$ $>$ 10\%). Other strong transitions  are to some 2$^{+}$, 3$^{+}$ and 4$^{+}$ levels and the 1679 keV resonance decays rather intensely also to the 3389 and 3756 keV 6$^{+}$ levels. The PC coefficients relating the decay of both resonances to the 1$^{+}$, 3$^{+}$ and 4$^{+}$ standad resonances are very close and indicate uniquely the value of the spin J~=~3. The $<$J$>$ test indicate J = 3 only for the 1687 keV resonance, while for the 1679 keV resonance rather the J = 4 value is expected, although J = 3 cannot be fully excluded.

All our results are in contradiction with assignments of the 1679 and 1687 keV resonances as the J$^{\pi}$ = 4$^{+}$ ones, done in ref. ~\cite{Chat}. We conclude from our analysis, that both resonances are the members of splitted J$^{\pi}$ = 3$^{+}$ isobaric analogue of the 341 keV state in $^{56}$Mn. Well observed decay to the 6$^{+}$ levels and the $<$J$>$ test indicate that in lower resonance participates substantially also other state with higher, probably J = 4 spin. The 3$^{+}$ 1696 keV resonance might belong to the isobaric multiplet, but absence of the transition to the 5132 keV state is in contradiction with such conclusion and the resonance should  be considered rather as independent one.

{\underline {\it The 1773 keV resonance.}} El-Kazass et al.~\cite{Kaz2}~interpreted the resonance at E$_{p}$~= 1766(4)~keV as the 4$^{+}$, 5$^{+}$ isobaric analogue of the 454 keV level in the $^{56}$Mn nucleus, but this level in $^{56}$Mn was later interpreted as pure 3$^{+}$ state~\cite{NDS}.

Radiative decay of resonance at E$_{p}$ = 1773 keV measured by us is in excellent agreement with ref. ~\cite{Chat}~ but only in a very rough agreement with Ref.~\cite{Kaz2}. Although the decay is strongly fragmented (I$_{\gamma}$ $>$ 1\% for 28 primaries) observed radiative strength S$_{\gamma}$ is only about 80\% of full strength S$_{0}$. Many intensive primaries to the 2$^{+}$, 3$^{+}$ and 4$^{+}$ levels of the $^{56}$Fe nucleus have been observed, the strongest ones exciting the 2085, 4298, 4510 keV  (4$^{+}$) and 2658 keV (2$^{+}$) levels (I$_{\gamma}$ = 10.3, 5.5, 5.5 and 4.4\%, respectively). According to both, $<$J$>$ and PC tests, spin of the resonance is J = 3, in discord with Ref. ~\cite{Kaz2}~. Remarkable lack of the radiative strength in comparison with the 1344 keV and other 3$^{+}$ resonances might indicate odd parity of the resonance but we have not found any other indication for this conclusion. Therefore although in table 1 the J$^{\pi}$ = 3$^{-}$ assignment for the resonance is included, the parity should be considered as very tentative.

{\underline {\it The 1807 keV resonance.}} Radiative decay of the resonance studied in present work, is in rough agreement with measurements ~\cite{Frid}~and ~\cite{Kaz1}. El-Kazass ~\cite{Kaz1}~interpreted this resonance as the 3$^{+}$ IA of the 486.3 keV 3$^{+}$ level in the $^{56}$Mn nucleus. Assignment is not in strong contradiction with our results. The $<$J$>$ test indicate J = 3 or higher while the PC-test relating decay of the 1807 keV resonance to other resonances is rather uncertain (all over PC $\sim$ 0.4 -- 0.7), although spin J = 3 is preferred. However, the PC values relating the resonance to all possible 3$^{+}$ resonances is only about 0.5 -- 0.6. 

The resonance decays very strongly to the 847 keV 2$^{+}$ level (I$_{\gamma}$ $\sim$ 37\%) but weak transitions to both 6$^{+}$ levels have been somewhere observed. On the other side, low fragmentation of the decay as well as rather low value of radiative strength (S$_{\gamma}$  $<$ 90\% of S$_{0}$) indicate higher spin of the resonance. Summing all information concerning the 1807 keV resonance, we cannot do any definite conclusion about its J$^{\pi}$ assignment. Therefore we assume that the resonance is basically the 3$^{+}$ IAR as assumed in ~\cite{Kaz1}, however with respect to the energy it corresponds to the 454 keV level in the $^{56}$Mn nucleus. Substantial participation of other compound nucleus states with spin J = 2 -- 4 in formation of the 1807 keV resonance seems to be very probable.\\

\subsubsection{Isobaric analogue resonances}

\begin{table}[b,t]
\centering
\begin{tabular}{|r|c|r|r|r|r|r|}
\hline
\multicolumn{2}{|c|}{Levels in $^{56}$Mn}
&$E_r^0$&   $E_p$& $E_r$&$<E_r>$&$\Delta E_{r}$\\
\cline{1-2}                       
$\varepsilon_{i}$[keV]&J$^{\pi}$&[keV]&[keV]&[keV]&[keV]&[keV]\\
\hline
\hline
      0    & 3+  &  11504  &   1344 &  11504  & 11504  &   0\\
\hline
           &     &         &   1441 &  11599  &        &    \\
    110.5  & 1+  &  11614  &   1455 &  11613  & 11605  &   9\\
           &     &         &   1446 &  11604  &        &    \\
\hline
           &     &         &   1521 &  11678  &        &    \\
    212.0  & 4+  &  11716  &   1524 &  11681  & 11692  &  34\\
           &     &         &   1531 &  11688  &        &    \\
\hline
    215.1 &1+,2+ &  11719  &   1536 &  11692  & 11692  &  27\\
\hline
    335.5  & 5+  &  11839  &   1727 &  11880  & 11883  & -44\\
           &     &         &   1734 &  11887  &        &    \\
\hline
           &     &         &   1679 &  11833  &        &    \\
    341.0  & 3+  &  11845  &   1687 &  11841  & 11841  &   4\\
           &     &         &   1696 &  11850  &        &    \\
\hline
    454.3  & 3+  &  11958  &   1773 &  11925 ?& 11925  &   33\\
\hline
    486.3  & 3+  &  11990  &   1807 &  11958 ?& 11958  &   32\\
\hline
\end{tabular}
\caption{Isobaric analogue resonances in the $^{56}$Fe nucleus}
\end{table}

Comparison of observed isobaric analogue resonances in the $^{56}$Fe nucleus with excited states in parent $^{56}$Mn nuclei is done in table 3 and it is based on the assumption that the E$_{r}$ = 11504 keV resonance is singlet and is the analogue of the ground state in $^{56}$Mn. Assuming that the energy spacing of the resonances should be given by spacing of excitation energy $\varepsilon$$_{i}$ of correspondig states in the parent $^{56}$Mn nucleus, we compare expected energy E$_{r}$ of the resonances calculated as E$_{r}^{0}$ = (11504 + $\varepsilon_{i}$) keV with observed values E$_{r}$. For known isobaric multiplets the value of E$_{r}$ is taken as the centre-of-mass energy of the multiplet, $E_{r} = <E_{r_{i}}>$. Difference of both values $\Delta E_{r} = E_{r}^{0} - <E_{r}>$ is in last column of table 3.

Except for the J$^{\pi}$ = 5$^{+}$ IAR both values of $E_r^0$ and $<E_r>$ are in rough agreement, although some systematic lowering of experimental values might be noted. Small differences between both energies could be a result of the $\Delta$T = 1 mixing. Assignment of the 1773 keV resonances as an IAR one is very doubtful (see part 4.1.3) and according to the energy as a more convenient candidate for the analogue of the 454 keV state in $^{56}$Mn can be the 1807 keV resonance. Analogue of the 486 keV state should be then out of measured proton beam region.

Assignment of the J$^{\pi}$ = 5$^{+}$ IAR in the $^{56}$Fe nucleus corresponding to the 335 keV state in $^{56}$Mn is very tentative (see part 4.1.2). Main argument for such assignment remains high spin of both resonances and lack of any other convenient resonances in this energy region, which might be the candidate for corresponding analogue state.

\section{Conclusion}

Very extensive experimental information about radiative decay of 21 resonances made it possible to study many properties of the $^{56}$Fe compound system at high excitation energy (E$_{x}$~$\sim$~11.5~--~12~MeV). Exact energy of measured primary $\gamma$-rays enabled us to determine very accurately the excitation energy of individual resonances and corresponding proton beam energy. Analysis of this energy pointed out splitting of some resonances. Measured decay branching to many low-lying bound states in the $^{56}$Fe nucleus was exploited for assignment of the spin-parity charakteristics of individual resonances. Similarity of some decays  made it possible to discuse isobaric analogue multiplets and their determination. Obtained results were used for short comparison of energy systematics of the IAR in the $^{56}$Fe and excited states in the $^{56}$Mn nuclei.

\end {document}